\documentclass[aps,prl,superscriptaddress,
twocolumn,
showpacs]{revtex4-1} 

\usepackage[colorlinks=true, linkcolor=blue, citecolor=blue, urlcolor=blue]{hyperref}
\usepackage{amsmath,amssymb,mathrsfs}
\usepackage{latexsym}
\usepackage{graphicx} 
\usepackage{epstopdf}
\usepackage{graphicx,epstopdf,color}
\usepackage{amsfonts}
\usepackage{bm}
\usepackage{bbm}
\usepackage{multirow}

\begin{document}



\title{Quasi isotropic underwater acoustic carpet cloak based on latticed pentamode metafluid}
\author{Zhaoyong Sun}%
\affiliation{Key Laboratory of Noise and Vibration Research, Institute of Acoustics, Chinese Academy of Sciences, 21 North 4th Ring Road, Beijing 100190, China}
\author{Xuecong Sun}%
\affiliation{Key Laboratory of Noise and Vibration Research, Institute of Acoustics, Chinese Academy of Sciences, 21 North 4th Ring Road, Beijing 100190, China}
\affiliation{School of Electronic, Electrical and Communication Engineering, University of Chinese Academy of Sciences, 19A Yuquan Road, Beijing, 100049, China}
\author{Han Jia}%
\email[Correspongding author:]{hjia@mail.ioa.ac.cn}
\affiliation{Key Laboratory of Noise and Vibration Research, Institute of Acoustics, Chinese Academy of Sciences, 21 North 4th Ring Road, Beijing 100190, China}
\affiliation{State Key Laboratory of Acoustics, Institute of Acoustics, Chinese Academy of Sciences, 21 North 4th Ring Road, Beijing, 100190, China}
\author{Yafeng Bi}%
\affiliation{Key Laboratory of Noise and Vibration Research, Institute of Acoustics, Chinese Academy of Sciences, 21 North 4th Ring Road, Beijing 100190, China}

\author{Jun Yang}%
\email[Correspongding author:]{jyang@mail.ioa.ac.cn}
\affiliation{Key Laboratory of Noise and Vibration Research, Institute of Acoustics, Chinese Academy of Sciences, 21 North 4th Ring Road, Beijing 100190, China}
\affiliation{School of Electronic, Electrical and Communication Engineering, University of Chinese Academy of Sciences, 19A Yuquan Road, Beijing, 100049, China}
\affiliation{State Key Laboratory of Acoustics, Institute of Acoustics, Chinese Academy of Sciences, 21 North 4th Ring Road, Beijing, 100190, China}
%
%
%
%
\date{\today}

\begin{abstract}
In this work, we present a practical design of quasi isotropic underwater acoustic carpet cloak with pentamode microstructure.
The quasi conformal transformation is not only used to obtain the required parameters, but also used to deform the retrieved regular pentamode material structure to the desired carpet, during which the effective parameters can be considered as not being affected too much. 
The nice and broadband cloak effect shows the success of the design.
This work will contribute to the experimental demonstration of pentamode acoustic carpet cloak.
Moreover, the technique in this work also can be used to design arbitrarily shaped devices with microstructures.
\end{abstract} 

\maketitle


\section{\label{sec:level1}Introduction}
Transformation method, which is used to design amazing devices such 
as cloaks for wave controlling, was initially proposed for electromagnetic waves\cite{pendry_controlling_2006,leonhardt_optical_2006}, 
and then generalized to acoustics soon\cite{cummer_one_2007,chen_acoustic_2007}.
It uses coordinate transformations to build connections between geometric space and material space.
In several fields, different cloaks, designed with the transformation method to cancel the scattering of the incident wave to hide objects, have been reported recently\cite{pendry_controlling_2006,
leonhardt_optical_2006,cummer_one_2007,zhang_broadband_2011,
chen_conformal_2011,gomory_experimental_2012,chen_latticed_2015,
chen_broadband_2017}.

Another most attracting application is the carpet cloak, which covers the object on a rigid plane and makes the scattering
wave perform as being reflected by the plane, resulting in the concealment of 
the object\cite{li_hiding_2008}.
The idea of carpet cloak was firstly proposed by Li and Pendry\cite{li_hiding_2008}, and then was fabricated for electromagnetic wave\cite{liu_broadband_2009,gabrielli_silicon_2009,valentine_optical_2009}.
The concept was extended to acoustics immediately\cite{pendry_acoustic_2008,zhu_numerical_2010}.
And soon latter, an acoustic carpet composed with perforated plates was designed by a linear transformation, where the materials are anisotropic but homogeneous\cite{popa_experimental_2011}.
The good cloak effect was shown by the experimental results. 
Since then, more relative researches on carpet cloak for airbone sound have been reported\cite{hu_experimental_2013,zigoneanu_three-dimensional_2014,esfahlani_acoustic_2016}.
Compared with carpet cloak for airbone sound, the underwater one has more important practical applications, but suffering from difficulty in impedance matching and microstructure design\cite{popa_homogeneous_2011,zhu_unidirectional_2015}.
In our previous works, we experimentally demonstrated the two and three dimensional under water carpets\cite{bi_design_2017,bi_experimental_2018}.
The carpets were constructed with uniformed mental strips alternating with water layers periodically,  which behave like fluid with anisotropic inertia and isotropic bulk modulus.
The effectiveness of the prototypes was experimentally demonstrated in anechoic tanks.
Another approach to under water carpet is using pentamode material (PM), which can be completely fabricated from rigid solids in practical case\cite{milton_which_1995}.
For the ideal PM in three dimensional (3D) case, the six order elastic tensor only has one nonzero eigenvalue. 
The five vanished eigenvalues correspond to five easyways to deform the materials, which leads to the zero shear modulus. 
This is the reason that the pentamode is so named (“penta” means “five” in Greek)\cite{milton_which_1995,norris_acoustic_2008}. 
The only eigenmode is a hydrostatic stress. 
Thus, the ideal PM exhibits similar acoustic properties to fluids that only the compressive wave can exist.
In two dimensional (2D) case, the name ``pentamode'' is stilled inherited, even though only two of the three eiganvalues are zeros.
The 2D PM can be realized by hexagonal lattice with highly tunable effective parameters, which can be designed to have anisotropic modulus\cite{kadic_practicability_2012,layman_highly_2013}.
The shear modulus of the latticed PM is so small that it can be neglected and regarded as fluid in some frequency bands.
PM has attracted great attention in underwater acoustic wave controlling\cite{norris_acoustic_2009,chen_latticed_2015,tian_broadband_2015,su_broadband_2017,chen_broadband_2017,sun_design_2018}.
It has been proved that wave equation in PM is also form-invariant under the coordinate transformations\cite{norris_acoustic_2008}, thus PM can be used to design some valuable under water devices with transformation method\cite{chen_latticed_2015,chen_broadband_2017}.
However, it is difficult to design an anisotropic PM carpet (PMC), because of the complexity of the transformation method and harsh boundary conditions of impedance matching\cite{norris_acoustic_2008,chen_design_2016}.
An possible way for PMC is quasi conformal transformation (QCT), which will bring the quasi isotropic parameter distribution.
The QCT method can be archived by grid deformation theory\cite{li_hiding_2008} or solving inverse Laplace's equations\cite{hu_design_2009,chang_design_2010}.
Although suffering from large size and  inhomogeneity, the arbitrary shapes, the minimized anisotropy and  broadband character of the QCT
based devices still show strong attractions\cite{li_hiding_2008,liu_broadband_2009,chang_design_2010,hu_design_2009}.

In this work, we use a 2D version of
the PM lattice to design a quasi isotropic latticed PMC.
The continuous required parameters are obtained with QCT method, which is archived by solving inverse Laplace's equations with Neuman and  Dirichelet boundaries\cite{hu_design_2009,chang_design_2010}.
In order to realize the carpet with 2D PM microstructures, the carpet is pre-divided into $300$ different cells, where the discrete parameters are obtained by the area average of the continuous one.
The 2D PM unit cells,  with effective parameters equaling to that of the corresponding discrete ones, are acquired by retrieving the energy bands\cite{chen_latticed_2015,sun_design_2018}. 
Since the retrieved PM unit cells are regular hexagon, it is difficult to build a carpet.
Thus, some appropriate deformations of the PM cells are necessary.
The QCT method is reapplied to deform the retrieved PM cells in order to build the desired carpet.
We firstly use the retrieved PM cells to form $300$ rectangle shaped bricks which have the same parameters with the discrete ones correspondingly, and then arrange them into an array having the same size with that of the virtual space.
The PM array is viewed as a new virtual space with the required parameters of the carpet, and mapped into a carpet with the same QCT.
Due to the conformal preservation, we suppose that the deformations will not change the effective parameters of the structure seriously.
This hypothesis is further supported by the good cloak effect of simulation results.
Cosine similarity is used to demonstrate the broadband and good performance of the latticed PMC in different directions.

\section{\label{sec: transformation}Design method}
\begin{figure*}[htp] 
\includegraphics[width=4.732in]{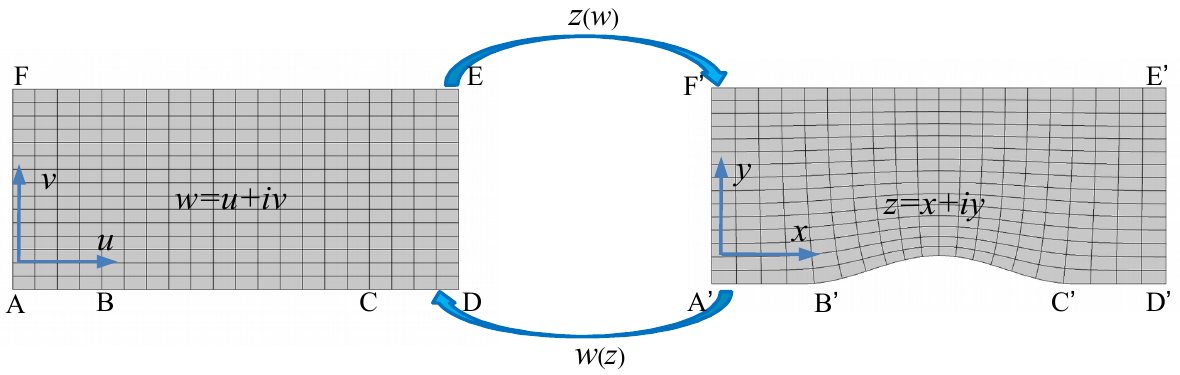}
\caption{\label{fig:fig1} (Color online) Illustration of conformal transformation acoustics. Conformal mapping $\omega (z)$ transforms the virtual space $ABCDEF$ into the physical space $A'B'C'D'E'F'$, while the inverse mapping $z(\omega)$ transforms the physical space back to the virtual space.}
\end{figure*}
Considering that the virtual space and physical space are described by complex coordinates $w=u+iv$ 
and $z=x+iy$ correspondingly,  the conformal transformation optics builds a relationship between those two spaces in the following way\cite{leonhardt_optical_2006,turpin_conformal_2010,xu_conformal_2014,xu_perfect_2016}:
\begin{equation} 
  n_z=n_w \lvert \frac{dw}{dz} \rvert =n_w \sqrt{\frac{1}{det J}} \label{eq:conf},
\end{equation}
where $n_z$ and $n_w$ denote the reflective index in physical and virtual spaces, and $J$ is the Jacobian transformation matrix. 
This result is derived by mapping the  Helmholtz equation in physical space into the virtual space with the analytical function $w(z)$\cite{leonhardt_optical_2006,xu_conformal_2014}.
Since the propagation of the acoustic wave in isotropic media also obeys the Helmholtz equation, the conformal transformation also can be used to design acoustic device.
Thus, the relationship in Eq.(\ref{eq:conf}) will still be kept in acoustics with conformal transformations. 

Conformal transformation provides the possibilities for isotropic carpet cloak designing.
Figure \ref{fig:fig1} shows the transformation of a carpet, where the virtual space $ABCDEF$ presents the free space with reflective index $n_w$ and the physical space $A'B'C'D'E'F'$ represents the carpet cloak filled with the transformed media with reflective index $n_z$.
The boundaries $A'B',C'D',E'F'$ and $D'E',F'A'$ are all straight lines along horizontal and vertical directions, while the arc $\widehat{B'C'}$ is an arbitrary curve.
The background media is water with $\rho_0=1000$ Kg/m$^3$, $K_0=2.25$ GPa, and $n_w=n_0=1$.
According to Eq.(\ref{eq:conf}), $n_z$ is determined by the transformation matrix $J$. 
However, finding a suitable mapping $z(w)$ for the designed carpet is a great challange, since analytical function $z(w)$ determines the shape of the physical space.
Quasi conformal transformation (QCT) is an appropriate option for this situation\cite{li_hiding_2008,hu_design_2009,chang_design_2010,ren_acoustic_2011}  
One way to realize QCT is calculating the inverse mapping $w(z)$ by numerically solving the inverse Laplace's equations with predefined boundary conditions\cite{hu_design_2009,chang_design_2010,ren_acoustic_2011}.

The QCT theory suggests that the Cauchy-Riemann conditions must be satisfied in the mapping $w(z)$:
\begin{subequations}
  \begin{align} 
    \frac{\partial u}{\partial x} &= \frac{\partial v}{\partial y} \\
    \frac{\partial u}{\partial y} &= - \frac{\partial v }{\partial x}
  \end{align}
  \label{eq:cr}
\end{subequations}
which can result to the Laplace equations as:
\begin{subequations}
  \begin{align}
   \nabla ^2 u& =0, \\
   \nabla ^2 v &=0.
 \end{align}
  \label{eq:Laplace}
\end{subequations}
According to the equivalence of the external boundaries between virtual and physical spaces, the Neuman and Dirichelet boundaries are used:
\begin{subequations}
\begin{align}
& \vec{n}\cdot \vec{\nabla}u(x,y)|_{A'B'C'D',E'F'}=0, & v(x,y)|_{A'B'C'D',E'F'}=y;\\
&\vec{ n}\cdot \vec{\nabla}v(x,y)|_{F'A',D'C'}=0, & u(x,y)|_{F'A', D'E'}=x.
\end{align} 
\label{eq:boundery}
\end{subequations}

Thus, the inverse Jacobian matrix $A=J^{-1}$, which is shown as follows: 
\begin{equation}
A=\left (
\begin{array}{cc}
    \frac{\partial u}{\partial x}&\frac{\partial u}{\partial y} \\
    \frac{\partial v}{\partial x}&\frac{\partial v}{\partial y}
\end{array}
\right ),
\end{equation}
can be obtained by solving the Laplace equation Eq.(\ref{eq:Laplace}) with the boundary Eq.(\ref{eq:boundery}) in a partial differential equation solver. 
Because the calculation is numerical, it is easy to be understood that $A_{11}=\partial u/ \partial x$ ($A_{12}=\partial u/\partial y$) does not exactly equal to $A_{22}=\partial v /\partial y$ ($-A_{21}=-\partial v /\partial x$).
This can lead some very soft anisotropy which can be neglected.
Thus, the transformation can be regard as quasi conformal.

If regarding the density as homogeneous, then the acoustic parameters can be written as:
\begin{subequations}
  \begin{align}
   \rho &= k \rho_0 ,\\
   K&=\rho c^2=\frac{k K_0}{n^2}=k K_0 det J,
  \end{align} 
  \label{eq:nor}
\end{subequations}
with $k >0$ as an arbitrary constant able to tune the parameters. 
Considering the identity $\lvert dw/dz \rvert =\sqrt{1/\lvert det J \rvert }$\cite{turpin_conformal_2010}, one can find that Eq.(\ref{eq:nor}) is the same with Norris's results\cite{norris_comment_2012}.

On the other hand, keeping the physical space with an impedance equaling to that of the background media, one can derive that the bulk modulus and density of the physical space have the properties shown as follows:
\begin{subequations}
\begin{align}
  &  \rho = \frac{\rho_0 c_0}{c}=\rho_0 n=\rho_0 \sqrt{\frac{1}{det J}} \label{eq:rho}\\
  &  K=\rho c^2=\frac{K_0}{n}=K_0 \sqrt{det J} \label{eq:K}
\end{align} 
\label{eq:cta}
\end{subequations}
which are the same with Eq.(6) in Ren $et ~  al$'s work\cite{ren_design_2010}.

In this paper, we chose the acoustic parameters shown in Eq.(\ref{eq:cta}) to design the latticed PMC. 
\section{\label{sec:unitcell}Design of the unit cells}
\begin{figure}[htp] 
  \includegraphics[width=3in]{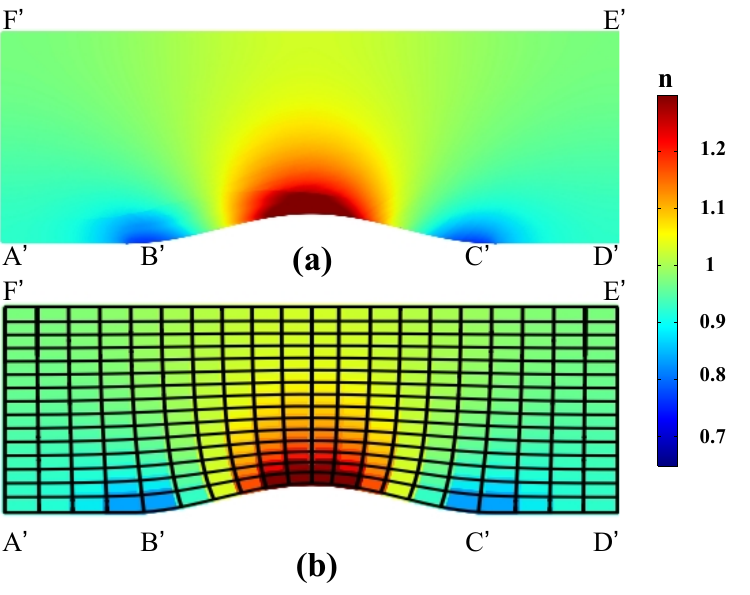}
  \caption{\label{fig:fig2}(Color online) Profiles of (a) the continuous refractive index and (b) the discrete approximation in the acoustic carpet. }
 \end{figure}
In our case, the designed carpet has the scale of $E'F'=4.8$ dm and $D'E'=2.0785$ dm, while the arc $\widehat{B'C'}$ is a curve described by $y=a\cos ^2(\pi x/b)$ with $a=0.3$ dm, $b=125/36$ dm, and $x\in [-1.44,  1.44]$ dm.
The background media is water with $\rho_0=1000$ Kg/m$^3$, $K_0=2.25$ GPa, and $n_w=n_0=1$.
The inverse Jacobian matrix $A$ is obtained by solving the Laplace equation Eq.(\ref{eq:Laplace}) with the boundaries Eq.(\ref{eq:boundery}) in a partial differential equation solver. 
Since the inverse Laplace's equations are numerically solved, the Cauchy-Riemann relationships shown in Eq.(ref{eq:cr}) are not exactly but approximately satisfied.
This causes  tiny anisotropy.
\begin{figure*}[htp]
  \includegraphics[width=6in]{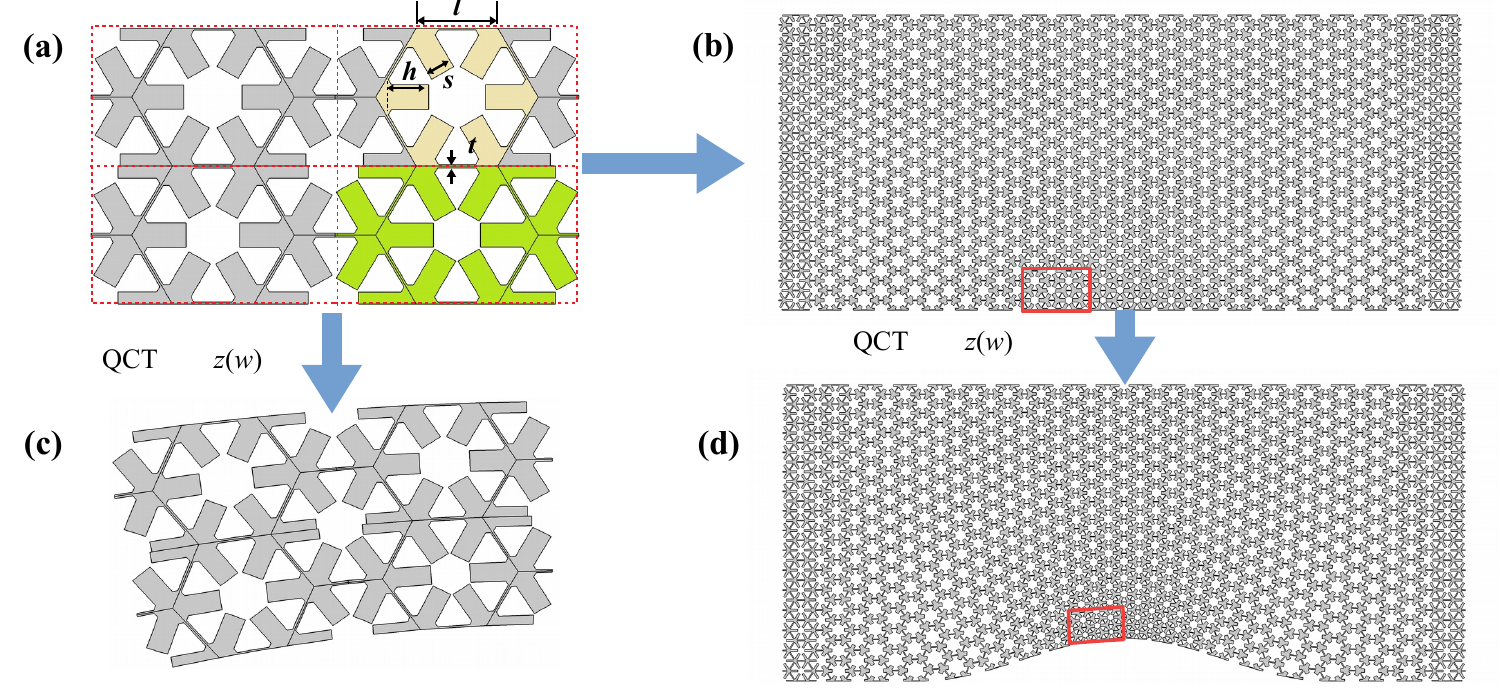}
  \caption{\label{fig:fig3}(Color online) (a) Microstructure of the 2D PM, with a basic PM unit cell colored in light yellow and a PM brick colored in lemon. (b) The $15\times 20 $ sized PM brick array  with  corresponding expected effective refractive index shown in Fig.\ref{fig:fig2}(b). (c) The deformed PM bricks archived by conformally transforming the regular PM bricks shown in (a). (d) The microstructure of the PMC transformed from the array shown in (b).}
 \end{figure*} 
The anisotropic ratio, defined as $A_{11}/A_{22}=\lvert A_{12}/A_{21}\rvert$ here,  is as small as $1.058$,  which can be neglected.
Then the inverse mapping $w(z)$ and the mapping $z(w)$ can be confirmed and regarded as nearly conformal.
The index profile in the acoustic carpet is shown in Fig.\ref{fig:fig2}(a), which presents a petal like distribution. 
 It tends to $n=1$, the value of water, near the boundary $D'E', E'F', F'A'$, and reaches at the maximum of $n=1.35$ at the peak of curve $\widehat{B'C'}$ and the minimum of $n=0.75$ near the points $B'$ and $C'$.
The carpet is pre-divided into $300$ different cells in order to be realized by PM structure.
The discrete indexes are shown as Fig.\ref{fig:fig2}(b).
There are $15$ rows and $20$ columns.
The index in each cell is the area average of the continuous one. 
Thus, it is easy to obtain the required bulk modulus and density of each cell according to Eq.(\ref{eq:cta}).
And in this work, only isotropic PM is in considerations.

A basic unit cell of PM is shown in Fig.\ref{fig:fig3}(a) which is colored in light yellow.
The substrate is aluminum with density $\rho_{Al}=2700$ kg/m$^3$, Young's modulus $E_{Al}=69$ GPa and Poisson's ratio $\nu=0.33$.
The regular hexagon cells behave as isotropic metafluid.
The geometric parameters of the basic PM unit cell are characterized by side length $l$, strut thickness $t$, and the block size $(s,h)$. 
At the quasi-static regime\cite{norris_mechanics_2014}, the effective density is the volume average of the mass.
Thus the effective density is mainly determined by the block size $(s,h)$.
The ratio of thickness and side length, $\eta=t/l$, mainly determines the effective bulk modulus of the PM.
The effective parameters of the PM can be calculated by Bloch-Floquet analysis\cite{kutsenko_wave_2017,sun_design_2018}, and the required geometric parameters can be obtained by retrieving the energy bands\cite{chen_latticed_2015,sun_design_2018}.

In order to construct the carpet with the latticed PM structures, the rectangle shaped PM brick is composed by one unit cell with four quarter cells adjoining the four oblique sides, shown as the lemon structure in Fig.\ref{fig:fig3}(a). 
And there are four different PM bricks marked by the red dashed lines in Fig.\ref{fig:fig3}(a).
Every brick, with the side length $l=8$ mm, has the same size with that of the grids in the virtual space shown in Fig.\ref{fig:fig1}.
All the bricks together pill up a $15 \times 20$ array (Fig.\ref{fig:fig3}(b)), where every PM brick has nearly the same effective index with the corresponding cells in Fig.\ref{fig:fig2}(b).
The red rectangle in Fig.\ref{fig:fig3}(b) marks the four PM bricks shown in Fig.\ref{fig:fig3}(a).
The PM bricks array has the same size with that of the virtual space.

Since the bricks are regularly shaped, some deformations should be implemented in order to form the designed carpet cloak.
We use the QCT technique again to deform the regular PM bricks into a appropriate shape which can form the needed carpet.
The PM brick array is regarded as a new virtual space, where the effective parameters are not the same with the back ground media any more.
Since the mapping $z(w)$ connects the virtual space and the physical space, $z(w)$ can transform the geometry of the PM array into a shape similar with the physical space.
Note that QCT can be regarded as approximate conformal transformation, thus it can keep the angles and shapes locally unchanged during the transformation. 
The basic assumption in the deformations is that  the transformation of the bricks can only change the scales and keep the shape almost unchanged.
As what have been mentioned above that the effective parameters of the latticed PM are determined by the shape not the size of the structures, the deformations will not change the effective parameters of the bricks seriously.
The formed carpet is shown in Fig.\ref{fig:fig3}(d).
It can be observed that most of the PM unit cells keep a highly approximate regular hexagon lattice.
We suppose that the deformed structure will keep the original effective parameters in the long wave approximation, which is also proved by the simulation results.
Since that the bottom boundary $\widehat{B'C'}$ is not specific, the technique can be used to carpet with arbitrarily shaped bottom boundary\cite{hu_design_2009,chang_design_2010}.

\section {simulation results}
\begin{figure*}[!htp]
  \includegraphics[width=6in]{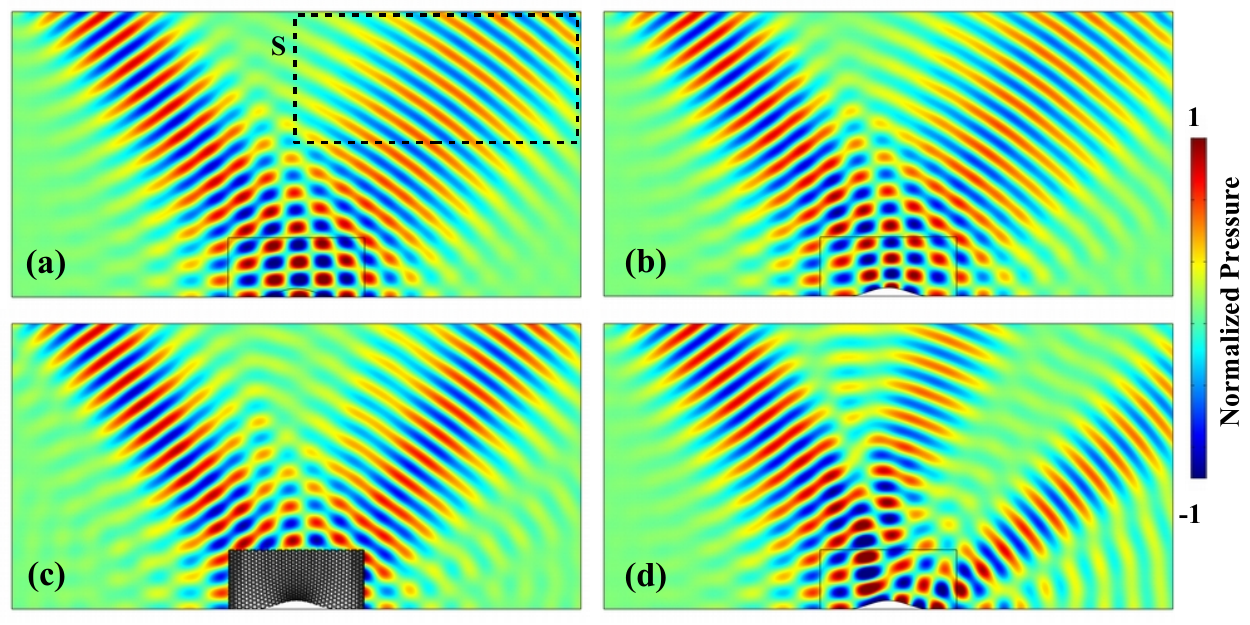}
\caption{\label{fig:fig4}(Color online) Acoustic pressure field of (a) rigid plane (b) ideal PMC (c) rigid scatter and (d) latticed PMC.}
 \end{figure*}

The performance of the latticed PMC is simulated using finite element solver COMSOL Multiphysics.
The simulated results are shown as Fig.\ref{fig:fig4}, where a Gassian beam of frequency $15$ Hz is incident at $55^\circ$
from the left up side to the bottom.

In Fig.\ref{fig:fig4}(a), the bottom side is set as acoustic rigid boundary to imitate the rigid ground plane.
It is obvious that all the incident waves are reflected by the rigid plane and the reflected waves keep the plane wave property.
Figure \ref{fig:fig4}(b) presents the acoustic field distribution of the ideal PMC with the index shown in Fig.\ref{fig:fig2}(a).
The field is almost identical to that of the rigid plane shown in Fig.\ref{fig:fig4}(a).
Figure \ref{fig:fig4}(c) shows the field distribution of latticed PMC, whose structure is presented in Fig.\ref{fig:fig3}(d).
In Fig.\ref{fig:fig4}(d), the incident beam is scattered into two distinct beams at different angles by the rigid bulge.
It shows that the reflected field is very similar to that of the ideal case in Fig.\ref{fig:fig4}(b).
Thus, it can be concluded that the designed carpet can cloak the scatter well at $15 kHz$.

\begin{figure}[!htp]
  \includegraphics[width=3in]{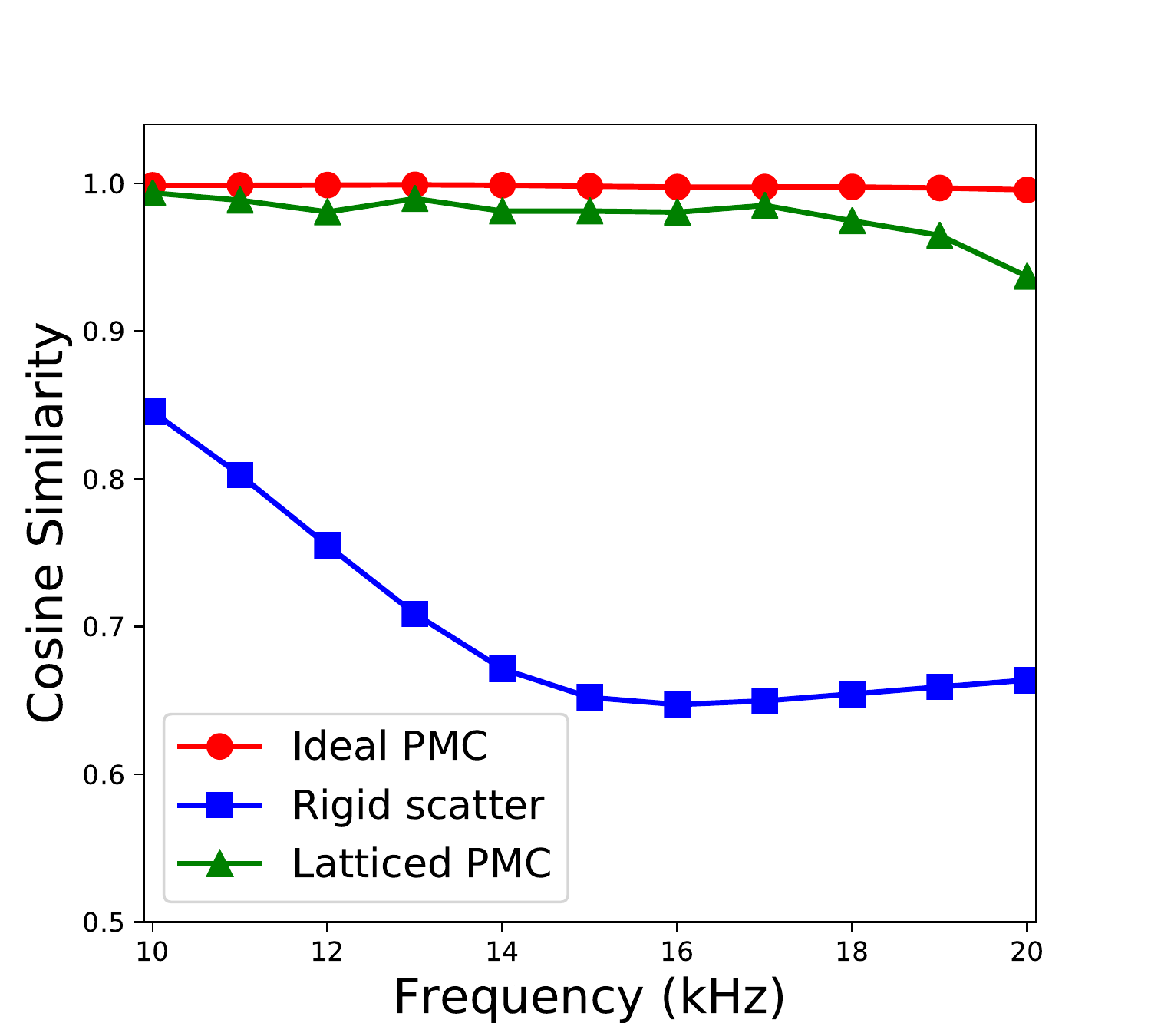}
\caption{\label{fig:fig5} (Color online) The cosine similarity of the reflected waves for ideal PMC (red circles), rigid scatter (blue squares) and latticed PMC (green triangles).}
\end{figure}
Moreover, the cosine similarity (CSI), which can be written as follows\cite{zhu_unidirectional_2015,xiong_design_2015}:
\begin{equation}
  CSI=\frac{P^rP^o}{|P^r||P^o|}=\frac{\Sigma P^r_i P^o_i}{\sqrt{\Sigma (P^r_i)^2}\sqrt{\Sigma (P^o_i)^2}},
  \label{eq:csi}
\end{equation} 
is used to evaluate the performance of the latticed PMC at different frequencies.
In Eq.(\ref{eq:csi}), $P^r$ and $P^o$ are the reference and objective pressure in a measured domain. 
The reference pressure in this work is the reflected pressure of the rigid bulge, while the objective one is that in measure.
The measure domain is a $10$ dm $\times $ $3$ dm rectangle that is $6$ dm above the bottom side, shown as the dashed line $S$ in Fig.\ref{fig:fig4}(a).
The value of CSI describes the degree of similarity between the objective field and the reference pressure, i.e., a larger SCI means a higher similarity.
Figure \ref{fig:fig5} shows the SCIs of the ideal PMC, rigid scatter and latticed PMC at the frequency ranging from $10$ Hz to $20$ Hz.
It can been seen that the CSIs of the ideal PMC are all $1$, which means the ideal PMC has an identical performance with the rigid plane.

The CSIs of the latticed PMC coincide with that of the ideal PMC very well, while 
the CSIs of the rigid scatter are much smaller than that of the ideal and latticed PMC.
Thus, the latticed PMC is as effective as the ideal one at a broadband frequency domain.
\begin{figure*}
\includegraphics[width=6in]{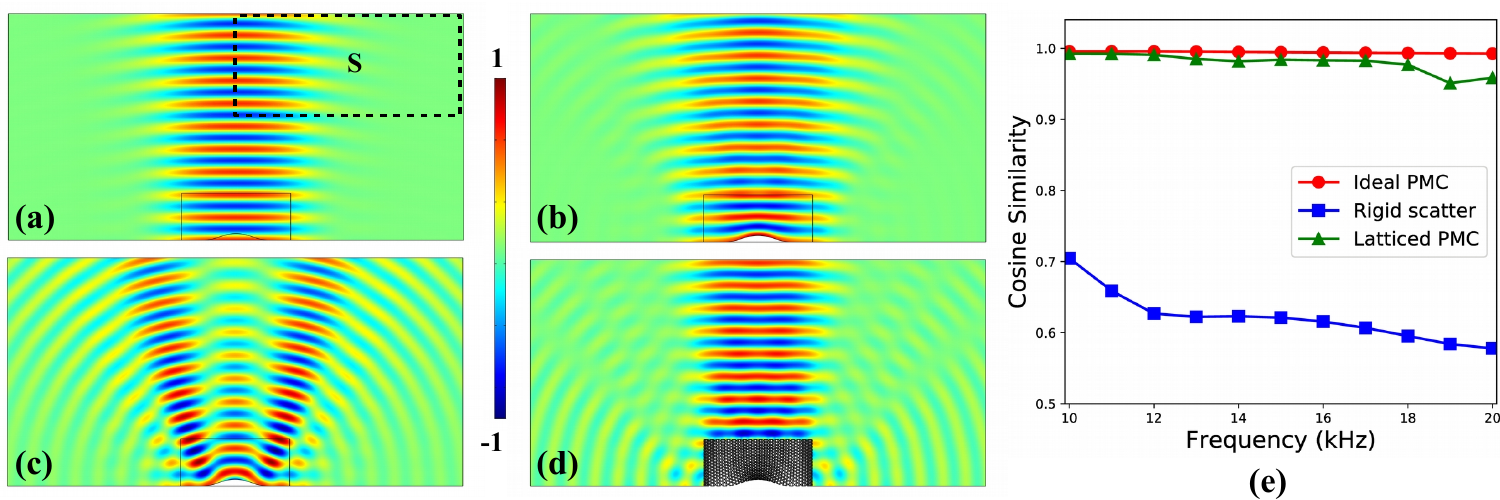}
\caption{\label{fig:fig6} (Color online) The scattered acoustic pressure of (a) the rigid plane, (b) the ideal PMC, (c) the rigid scatter and (d) the latticed PMC at $15$ kHz. (e) The cosine similarity of the scattered pressure for ideal PMC (red circles), rigid scatter (blue squares) and latticed PMC (green triangles)}
\end{figure*}

Furthermore, the case of the incoming beam incident in vertical direction is also studied, and the simulation results are shown in Fig.\ref{fig:fig6}.
It is obvious that in vertical incident case, the latticed PMC exhibits the same good cloak behavior with the oblique incidence.

The almost perfect performance of the ideal PMC indicates that the QCT can be used to design acoustic device.
Also, the good cloak effect of the latticed PMC shows that the deformation discussed in Sec.\ref{sec:unitcell} does not impact the effective parameters seriously.
\section{\label{sec:conslusion}Conclusion}

We have designed a latticed PM carpet via the QCT, which only needs an isotropic reflective index distribution.
The Jacobian transformation matrix, which determines the index distribution, is obtained by numerically solving the Laplace's equations in the inverse mapping.
The needed density and bulk modulus are calculated by keeping the impedance equal to that of water.
The designed carpet are divided  into $300$ cells in order to realize the latticed PMC by the 2D PM unit cells.
The corresponding $300$ PM bricks are obtained and arranged into an array with the same size with that of the virtual space.
We use QCT again to map the geometry of the PM brick array, which is regarded as a new virtual space, into a carpet shaped structure, in which process we assume the transformation will not change the effective parameters of the corresponding PM lattices too much.
The CSI in a chosen domain is calculated  to archive an accurate analysis for the cloak effect, which shows that both the ideal and latticed PMC have a highly similar behavior with the rigid plane.
This confirms the validation of the QCT method on carpet designing and unit cell deformations.
We hope the technique used in this work can be used to design relative isotropic acoustic devices.

\section{ACKNOWLEDGMENT}
\begin{acknowledgments}
The authors sincerely acknowledge the financial support of the National Natural Science Foundation of China (Grant No. 11874383), the Youth Innovation Promotion Association CAS (Grant No. 2017029) and the IACAS Young Elite Researcher Project (Grant No. QNYC201719).
\end{acknowledgments}

  
\providecommand{\noopsort}[1]{}\providecommand{\singleletter}[1]{#1}%

\end{document}